\documentclass[preprint,showpacs,preprintnumbers,amsmath,amssymb]{revtex4}

\usepackage{graphicx}% Include figure files
\usepackage{dcolumn}% Align table columns on decimal point
\usepackage{bm}% bold math
\usepackage{amsmath,dsfont}
\usepackage{amssymb,amsthm,amscd, amsbsy, array}

\newcommand{\half}{{\scriptstyle{\frac{1}{2}}}}
\def\2{{\half}}
\newcommand{\const}{\mathop{\rm const}\nolimits}
\def\vPi{{\bm{\Pi}}}
\def\va{{\bm{a}}}
\def\vc{{\bm{c}}}
\def\vb{{\bm{b}}}

\newcommand{\vA}{{\bm A}}
\newcommand{\vB}{{\bm{B}}}

\newcommand{\vP}{{\bm P}}
\newcommand{\vQ}{{\bm Q}}
\newcommand{\vq}{{\bm q}}
\newcommand{\vp}{{\bm p}}
\def\vE{{\bm E}}

\def\beq{\begin{equation}}
\def\eeq{\end{equation}}
\def\beqa{\begin{eqnarray}}
\def\eeqa{\end{eqnarray}}
\def\nn{\nonumber}
\def\barray{\left(\begin{array}}
\def\earray{\end{array}\right)}
\def\barraynb{\begin{array}}
\def\earraynb{\end{array}}

\def\IR{{\mathds{R}}} %%%%% Reals

\def\SO{{\rm SO}}
\def\smallover#1/#2{\hbox{$\textstyle\frac{#1}{#2}$}} %

\def\vx{{\bm{x}}}

\def\vu{{\bm{u}}}
\def\vX{{\bm{X}}}

\def\cK{{\cal K}}
\def\cP{{\cal P}}
\newcommand{\vcK}{{\bm \cK}}
\newcommand{\vK}{{\bm K}}

\begin{document}

\preprint{arXiv: 1111.1595v2 [hep-th]}

\title{Kohn condition and exotic Newton-Hooke symmetry in the non-commutative Landau problem}

\author{P-M. Zhang}\email{zhpm@impcas.ac.cn}

\author{P.~A.~Horvathy\footnote{Permanent address: 
{\it Laboratoire de Math\'ematiques et de Physique
Th\'eorique},
 Universit\'e de Tours 
(France).}}\email{horvathy@lmpt.univ-tours.fr}
\affiliation{Institute of Modern Physics, Chinese Academy of Sciences
\\
Lanzhou (China)
}

\date{\today}

\begin{abstract}
$N$  ``exotic'' [alias non-commutative] particles with 
masses $m_a$, charges $e_a$ and non-commutative parameters
$\theta_a$,  moving
in a uniform magnetic field $B$, separate into center-of-mass and
internal motions if Kohn's condition $e_a/m_a=\const$
is supplemented with $e_a\theta_a=\const.$
Then the center-of-mass  behaves as a single exotic particle carrying the total mass and
charge of the system, $M$ and  $e$, 
 and  a suitably defined non-commutative parameter $\Theta$. 
For vanishing electric field
off the critical case $e\Theta B\neq1$, 
the particles perform the usual cyclotronic motion with modified but equal frequency.
The system is symmetric under suitable time-dependent translations  which span a $(4+2)$-parameter centrally extended subgroup of the ``exotic'' [i.e., two-parameter centrally extended] Newton-Hooke group. In the critical case $B=B_c=(e\Theta)^{-1}$ the system is 
frozen into a static ``crystal'' configuration. Adding a constant electric field, all particles perform, collectively, a cyclotronic motion 
combined with a drift perpendicular to the electric field when $e\Theta B\neq1$. For $B=B_c$ the cyclotronic motion is eliminated and all particles move, collectively, following the Hall law. Our time-dependent symmetries
are reduced to the (2+1)-parameter Heisenberg group of centrally-extended translations.
\end{abstract}

%\\[8pt]
%Physics Letter {\bf B } (2012)  
%}

%ZH-Kohn-II-PLB-corr.tex arXiv: 1111.1595v2 in hep-th

\maketitle

%%%%%%%%%%%%%%%%%%%%%%
\section{Introduction}
%%%%%%%%%%%%%%%%%%%%%%

Kohn's theorem \cite{Kohn} says that  
a system of charged particles 
in a uniform magnetic field can be decomposed into center-of-mass and relative coordinates
if the charge-to-mass ratios are the same for all particles,
\beq
\frac{e_a}{m_a}=\const =\frac{e}{M}\,,
\label{KohnCond}
\eeq
where $e=\sum_ae_a$ and $M=\sum_am_a$ are the total charge and mass, respectively.
Renewed interest  in the question \cite{GiPo,ZH-Kohn}
comes from its relation 
 to Newton-Hooke symmetry \cite{NHlit,Arratia,HHimport}.
 
For an isolated system, the possibility of having a center-of-mass decomposition
relies on the
non-trivial cohomology of the Galilei group \cite{SSD}. In
 $d\geq3$ space dimensions the latter is the
 same as the one which generates
the central extension by the mass. 
In the plane the Galilei  group also admits
an ``exotic'' central extension  \cite{otherNC,DHexo}, though,
 highlighted by the non-commutation of boosts.  
 ``Exotic'' Galilean symmetry is realized by
non-commutative particles in the plane;
such a particle can be coupled to an  electromagnetic field, leading to 
the non-commutative Landau problem \cite{NCLandau}. 
In the critical case $B=B_c=(e\Theta)^{-1}$  the system becomes singular, and all motions follow a
generalized Hall law \cite{DHexo,NCLandau,Kokado}.

In this Note we combine and extend these results 
 to a system of $N$ exotic particles in the plane.
First we briefly review some aspects of the Landau problem for $N$ ordinary particles.

Our new results are presented in Sections \ref{Exotic} and \ref{Hall}~: 
we first generalize Kohn's theorem to exotic particles, and then we 
establish their symmetry under suitable time-dependent translations.  Off the critical case i.e. for $e\Theta B\neq1$ our
 conserved quantities realize  the
two-fold ``exotic'' extension  of
the Newton-Hooke group \cite{Arratia,AGKP}, further extended,
as in the Galilean case \cite{SSD,ZH-Kohn}, by internal rotations and time translations.

In Section \ref{Hall} we show how the Hall effect arises.
In the critical case $B=B_c$ all particles
move, collectively, as dictated by the Hall law.
The dimension of the phase space of the system drops from $4N$ to $2N$, and our ``time-dependent
translation'' -- symmetry reduces to the \emph{Heisenberg group} with $-(\Theta)^{-1}=-eB_c$ as central  parameter.

We find it worth noting that, in both the regular and singular cases, the motion is fully determined by the respective conserved quantities.

%\goodbreak
%%%%%%%%%%%%%%%%%%%%%%%%%%%%%%%%%%%%%%%%%%%%%%%%%%%%%%%%%%%%%%%%%%%%%%%%%%%%%%%%%%%%%%%%%%%%%%%%%%%%%%%%%%%%%%%%
\section{Kohn's theorem and  Newton-Hooke symmetry of the Landau problem}\label{Ordinary}
%%%%%%%%%%%%%%%%%%%%%%%%%%%%%%%%%%%%%%%%%%%%%%%%%%%%%%%%

A simplest proof of Kohn's theorem is obtained
by  using the equations
of motion; alternatively, the standard Lagrangian
can be decomposed 
as $L=L_{com}+L_{int}$,
where  $L_{com}$ and $L_{int}$ only depend on the center-of-mass,
$\vX=\sum_am_ax_a/M$ and
 the relative coordinates, $\vx_{ab}=\vx_a-\vx_b$,
 respectively \cite{GiPo}.
Recent interest in Kohn's theorem arose 
\cite{GiPo,ZH-Kohn} when it was noticed that 
when (\ref{KohnCond}) holds, the ``time-dependent translation'' (or ``boost''),
\beq
\vx_a\to \vx_a+\vb(t)
\label{btboost}
\eeq
is a \emph{symmetry} of the equations of motion
whenever  $\vb$ satisfies 
\beq
 M\ddot{\vb}=e\,\dot{\vb}\times\vB.
\label{NH}
\eeq 
  (\ref{btboost}) plainly leaves the relative coordinates 
invariant and only act on  the center-of-mass. The Lagrangian changes, moreover, as
\beqa
L\to L
-\big(\vX+\frac{1}{2}\vb\big)\cdot\underbrace{
\Big\{M\ddot{\vb}-e\,\dot{\vb}\times\vB\Big\}}_{0}
+\,
\frac{\ d}{dt}
\left\{\big(\vX+\frac{1}{2}\vb\big)\cdot
\big(M\dot{\vb}+\frac{e}{2}\,\vb\times\vB\big)
\right\},
\label{LBchange}
\eeqa
confirming that (\ref{btboost}) is indeed a symmetry
when (\ref{NH}) holds.
Working in the plane,  this equation  is solved by
$ 
\vb(t)=
R(-\omega t)\,
\va+\vc\,,
$ 
where $\va$ and $\vc$ are constant  vectors, $\omega=eB/M$, and $R$ denotes a planar rotation.  For $\va=0$ we get an ordinary translation and for 
$\vc=0$  a ``boost'' \footnote{They are  in fact``imported Galilean translations and 
boosts" \cite{ZH-Kohn, HHimport}.}. Adding rotations and  time translations provides us with \emph{Newton-Hooke transformations} \cite{NHlit,Arratia,AGKP}. 
 Then Noether's theorem provides us  with
 conserved ``magnetic momentum'' and 
``magnetic center-of-mass'', 
 \beqa
\Pi_i=M\big(\dot{X_i}-\omega\varepsilon_{ij}X_j\big),
\qquad
\vK=MR(\omega t)\dot{\vX},
\label{magmomboost}
\eeqa
 respectively. Eliminating $\dot{\vX}$ yields the usual cyclotronic motion,  
\beq 
X^i(t)=\frac{\varepsilon_{ij}}{eB}
\left(\Pi^j-\big(R(-\omega t)\vK\big)^j\right).
\eeq 
The value of $\vPi$ determines  the center around which the vector $\vK$ rotates with  common frequency $\omega$, shared by all particles.

%%%%%%%%%%%%%%%%%%%%%%%%%%%%%%%%%%%%%%%%%%%%%%%%%%%%%%%%%%%
%\subsection{Hamiltonian framework and Newton-Hooke Symmetry} 
%%%%%%%%%%%%%%%%%%%%%%%%%%%%%%%%%%%%%%%%%%%%%%%%%%%%%%%%%%%

Moreover, the usual one-particle commutation relations
$
\{p_a^i,p_b^j\}=e_aB\varepsilon^{ij}\delta_{ab},\,
\{x_a^i,p_b^j\}=\delta^{ij}\delta_{ab},\,
\{x_a^i,x_b^j\}=0,
$ imply that 
\beq
\{P^i,P^j\}=eB\varepsilon^{ij},
\quad
\{X^i,X^j\}=0,
\quad
\{X^i,P^j\}=\delta^{ij},
\eeq
where $\vP=\sum_a\vp_a$.  The 
conserved quantities  (\ref{magmomboost}) satisfy therefore
\beqa
\{\Pi^i,\Pi^j\}=-M\omega\,
\varepsilon^{ij},
\quad
\{K^i,K^j\}=M\omega\,\varepsilon^{ij},
\quad
\{\Pi^i,K^j\}=0.
\label{CMcomrel}
\eeqa 
These relations are consistent with those of the  mass-centrally-extended Newton-Hooke group  \cite{GiPo,ZH-Kohn,NHlit,Arratia}
\footnote{The same symmetry was found earlier for
Chern-Simons vortices in a constant electromagnetic field, see Ref. \cite{HHimport}.}. 
We just mention that
the equations of motion [omitted here] are  Hamilton's equations for $h =\sum{\vp_a^2}/{2m_a}$;
splitting $h$ as  $h=H+h_{int}$, where
\beq
H=\frac{\vP^2}{2M},
\qquad
h_{int}=
\frac{1}{4}\sum_{a\neq b}\frac{m_am_b}{M^3}\vp_{ab}^2,
\quad
\vp_{ab}=\frac{M}{m_am_b}(m_b\vp_a-m_a\vp_b)
\label{hamDecomp}
\eeq
shows that $H$ alone 
would yield the center-of-mass equations, owing to
 $\{X^i,h_{int}\}=\{P^i,h_{int}\}=0$.  

%%%%%%%%%%%%%%%%%%%%%%%%%%%%%%%%%%%%%%%%%%%%%%%%%%%%
\section{The Landau problem for exotic particles}\label{Exotic}
%%%%%%%%%%%%%%%%%%%%%%%%%%%%%%%%%%%%%%%%%%%%%%%%%%%%

Let us now consider $N$ ``exotic'' particles endowed with
masses, charges and non-commutative parameters
$m_a,\,e_a$ and $\theta_a$, respectively,  moving in a planar electromagnetic field. Although our theory works for any $B$ and $\vE$, we assume, for
simplicity,  that both fields  are constant.
Generalizing the $1$-particle equations in \cite{DHexo}, we describe our system by
\beqa
m^*_a
\dot{x}^i_a=p^i_a-m_ae_a\theta_a\varepsilon^{ij}E^j,
\qquad
\dot{p}^i_a=e_aB\varepsilon^{ij}\dot{x}^j_a
+e_aE^i,
\label{exoNeqmot}
\eeqa
where $m^*_a=m_a(1-e_a\theta_a B)$ is the effective mass
of the particle labeled by $a=1,\dots,N$.
Note, in the first relations,
 also  the ``anomalous velocity terms''
perpendicular to the electric field.
Summing over all particles, we find that when
${e_a}/{m_a}$ and $e_a\theta_a$ are both constants
i.e. when the \emph{generalized Kohn conditions} 
\beq
\frac{e_a}{m_a}=\frac{e}{M},
\qquad
e_a\theta_a=e\,\Theta,
\qquad
\Theta=\frac{\sum_am_a^2\theta_a}{M^2}  
\label{exoKohnCond}
\eeq
 hold (where
$e=\sum_ae_a$ is the total charge),
then the center-of-mass splits off,
\beqa
M^*\dot{X}^i=P^i-M{e}\Theta\varepsilon^{ij}E^j,
\quad
\dot{P}^i=eB\varepsilon^{ij}\dot{X}^j+
eE^i,
\quad
M^*=M(1-{e}\Theta B).
\label{exoCMeqmot}
\eeqa
The center-of-mass behaves  hence as 
a \emph{single ``exotic'' particle} carrying the
total mass, charge and non-commutative parameter, 
$M,\,e$ and $\Theta$, respectively. 

We first consider the purely magnetic case $\vE=0$, postponing that of $\vE\neq0$ to Sect. \ref{Hall}. Then  the center-of-mass performs the usual cyclotronic motion
but with modified frequency,
\beq
\vX(t)=R(-\omega^*t)\vA+\vB,
\qquad
\omega^*=\frac{eB}{M^*}=\frac{\omega}{1-e\Theta B }\,,
\label{exocyc}
\eeq
where $\vA$ and $\vB$ are constant vectors.

Now we  generalize the  symmetry
(\ref{btboost}). Ignoring the relative coordinates, 
 an easy calculation shows 
that a time-dependent boost (\ref{btboost}), implemented as
$
\vX\to\vX+\vb,\,
\vP\to\vP+M^*\dot{\vb},
$ 
 is a \emph{symmetry} for (\ref{exoCMeqmot}) whenever $\vb$ satisfies
\beq
M^*\ddot{b}^i-eB\varepsilon^{ij}\dot{b}^j=0.
\label{exoNH}
\eeq
cf. (\ref{NH}).
Alternatively, we note that the $N$-body exotic equations (\ref{exoNeqmot}) derive from the first-order phase space Lagrangian \cite{DHexo},
\beq
L_{exo}=\sum_a\left\{(\vp_a+e_a\vA_a)\cdot\dot{\vx}_a
-\frac{\vp^2_a}{2m_a}
+\frac{\theta_a}{2}\vp_a\times d\dot{\vp}_a\right\}.
\label{exoLag}
\eeq
When the generalized Kohn conditions (\ref{exoKohnCond}) hold, our Lagrangian can be decomposed into the sum of center-of-mass
and internal parts, $L_{exo}=L_{com}+L_{int}$, with \footnote{An interaction potential $V(\vx_a-\vx_b)$  would modify the equations of motion
(\ref{exoNeqmot}) by contributing to the electric
fields felt by each particle. The extra
terms would, however, drop out under summing, leaving
the center-of mass equations unchanged and affecting 
 the internal motion only. Since our main interest lies in the center-of-mass motion, 
we only consider $V=0$  in what follows.},
\begin{equation}
L_{com}=\left(\vP+e\vA\right) \cdot \dot{\vX}-\frac{\vP^2}{2M}+\frac{\Theta}{2}\vP\times\dot{\vP},
\label{exoCMLag}
\end{equation}
\begin{eqnarray}
L_{int} &=&\frac{1}{2}\sum_{a\neq b}
\frac{m_am_b}{M^2}\big(\vp_{ab}
+e\vA_{ab}\big)\cdot\dot{\vx}_{ab}
-h_{int}
+\frac{\Theta}{4}\sum_{a\neq b}\big(\frac{m_am_b}{M}\big)^2\vp_{ab}\times\dot{\vp}_{ab}\,,\qquad
\end{eqnarray}
where $\vA_{ab}=\vA_{a}-\vA_{a}$.

Let us now consider a time-dependent
translation $\vb(t)$. Implementing it on  $\vp_a$ according to
$\vp_a\to\vp_a+m_a^*\dot{\vb}$, we see that it leaves invariant the internal part, and only acts (consistently with our earlier implementation) on the center-of-mass.
Then a tedious calculation shows that our Lagrangian changes 
as
\begin{eqnarray}
\delta L_{exo} &=&\frac{d}{dt}\left[ \frac{1}{2}\Theta M^{*}\varepsilon^{ij}P^j\dot{b}%
^i+\frac 12M^{*}b^i\dot{b}^i+M^{*}\dot{b}^iX^i+\frac 12eB\varepsilon^{ij}b^iX^j\right]   
\nonumber \\[6pt]
&&+\left(\frac{\Theta}{2} P^i\varepsilon^{ij}-\frac 12b^j-X^j\right) \left(
M^{*}\ddot{b}^j-eB\varepsilon ^{jk}\dot{b}^k\right),
\label{NCquasiinv}
\end{eqnarray}
which is a surface term when  (\ref{exoNH})
is satisfied.

Assume first that we are off the critical case, 
$M^*\neq0$, i.e., $e\Theta B \neq1$. Then (\ref{exoNH})
is solved as in the ordinary case but with  frequency 
$\omega\to\omega^*$, 
\beq
\vb(t)=
R(-\omega^*t)\,
\va+\vc\,,
\label{exovasol}
\eeq
where $\va$ and $\vc$ are constant  vectors.
 The associated conserved quantities are therefore
\beqa
\cP^i&=&M^*\big(\dot{X^i}-\omega^*\varepsilon^{ij}X^j\big),
\\[6pt]
\vcK&=&\frac{M^*}{M\,\, \,}\,R(\omega^* t)\vP
=\frac{{M^*}^2}{M\,\,}\,R(\omega^* t)\dot{\vX}.
\label{exomagmomboost}
\eeqa
Eliminating $\dot{\vX}$ yields the classical trajectories
cf.  (\ref{exocyc}),
\beq 
X^i(t)=\frac{\varepsilon_{ij}}{eB}
\left(\cP^j-\frac{\big[R(-\omega^* t)\vcK\big]^j}{1-e\Theta B}\right).
\label{exotraject}
\eeq

%%%%%%%%%%%%%%%%%%%%%%%%%%%%%%%%%%%%%%%%%%
%\subsection{Exotic Hamiltonian structure}
%%%%%%%%%%%%%%%%%%%%%%%%%%%%%%%%%%%%%%%%%%

The  commutation relations of our system of exotic particles \cite{DHexo,NCLandau},
\beq
\{p_a^i,p_b^j\}=\frac{e_aB\varepsilon^{ij}}{1-e_a\theta_aB}
\delta_{ab},
\quad
\{x_a^i,p_b^j\}=\frac{\delta^{ij}}{1-e_a\theta_a B}
\delta_{ab},
\quad
\{x_a^i,x_b^j\}=\frac{\theta_a\varepsilon^{ij}}{1-e_a\theta_a B}
\delta_{ab},
\eeq
imply the center-of-mass Poisson brackets,
\beq\barraynb{lll}
\{P^i,P^j\}&=&\displaystyle\frac{eB\varepsilon^{ij}}{1-{e}\Theta B}\,,
\\[4pt]
\{X^i,P^j\}&=&\displaystyle\frac{\delta^{ij}}{1-{e}\Theta B}\,,
\\[4pt]
\{X^i,X^j\}&=&\displaystyle\frac{\Theta\varepsilon_{ij}}{1-{e}\Theta B}\,.
\earraynb
\label{exoCMcommrel}
\eeq
These Hamiltonian structures would allow us to recover both
exotic equations of motion (\ref{exoNeqmot}) and
 (\ref{exoCMeqmot}).
Moreover, the commutation relations of the exotic conserved quantities (\ref{exomagmomboost}) are,
\beqa
\{\cP^i,\cP^j\}=-M^*\omega^*\varepsilon^{ij}, 
\quad
\{\cK^i,\cK^j\}=(1-e\Theta B )M^*\omega^*\varepsilon^{ij},
\quad
\{\cP^i,\cK^j\}=0.
\label{exoCMcomrel}
\eeqa
Our new formulae can be compared to those in the ordinary i.e.  case $\theta_a=0$ in eqn. (\ref{CMcomrel}). Remembering that $M^*\omega^*=M\omega=eB$ we see that the commutation relations of the conserved magnetic momenta $\cP^i$
are the same as  in the commutative Landau problem.
The boost-with-boost relation are also similar to those in 
(\ref{CMcomrel}) to which they reduce when $\Theta=0$. For
$\Theta\neq0$, however, 
they pick up the characteristic factor $1-e\Theta B $.
The new relations (\ref{exoCMcomrel}) correspond in fact to 
 the   \emph{two-parameter centrally-extended ``\emph{exotic}''
Newton-Hooke symmetry} $\widetilde{NH}$ \cite{Arratia,AGKP}.
We only  mention here that, like in the Galilean case \cite{DHexo,NCLandau}, the non-commutativity  contributes to the 
c-o-m angular momentum, 
\beq
J=\vX\times\big({\vP}+eA(\vX)\big)+\frac{\Theta}{2}\vP^2,
\eeq
where the ``exotic'' contribution, 
$(\Theta/2)\vP^2=M\Theta H$, comes from the
manifestly rotation-invariant exotic term in
(\ref{exoCMLag}).
Moreover, if the interaction potential between the particles is itself
radial, $V=V(|\vx_a-\vx_b|)$,
then the system will be also invariant w.r.t.
internal rotation around the center-of-mass,
inducing a second, conserved, \emph{``internal angular momentum''},
\beqa
j_{int}&=&\frac{1}{2}\sum_{a,b}\frac{m_am_b}{M^2}(\vx_{ab})\times\left(
\vp_{ab}+eA_{a}\right)+M\Theta h_{int}.
\eeqa
The  ``exotic''contribution here is again proportional to the internal energy, $h_{int}$ in (\ref{hamDecomp}).
All this is understood by observing
that, for $V=0$, the equations of internal motions,
\beq
M^*\dot{\vx}_{ab}=\vp_{ab}\,,
\qquad
\dot{p}^i_{ab}=eB\varepsilon^{ij}\dot{x}_{ab}^j\,,
\label{inteqmot}
\eeq
have indeed the same form as those of the center of motion, (\ref{exoCMeqmot}).
The internal motions are therefore once again cyclotronic with   common frequency $\omega^*=eB/M^*$. 

The two terms
in the total  angular momentum,  $j=J+j_{int}$, are separately conserved 
-- as are $H$ and $h_{int}$ in (\ref{hamDecomp}).
The latter are in fact associated 
with ``internal rotations'' and ``internal time translations''
which act as symmetries, \emph{independently} of those acting on the center of mass. 
In conclusion, for $N\geq2$, the full
symmetry of the system is
\beq
\widetilde{NH}\otimes\big(\SO(2)\times\IR\big),
\label{totalsymmetry}
\eeq
in analogy with the Galilean, and extending the commutative cases \cite{SSD,ZH-Kohn}.

%%%%%%%%%%%%%%%%%%%%%%%%%%%%%%%%%%%%%%%%%%%%%%%%%%%
\section{The Hall effect in non-commutative mechanics}\label{Hall}
%%%%%%%%%%%%%%%%%%%%%%%%%%%%%%%%%%%%%%%%%%%%%%%%%%%

The most interesting physical application of the exotic model is to the Hall effect  \cite{DHexo,NCLandau,Kokado,QHE} that we now extend
to $N$ exotic particles. We first assume that the
electric field is turned off. Then,
in the critical case, 
\beq
M^*=0\quad\hbox{i.e.}\quad
B=B_{c}=\frac{1}{e\Theta},
\label{critcase}
\eeq
the only way to satisfy
the equations of motion (\ref{exoCMeqmot}) is
$\vP=\dot{\vX}=0$ and thus $\vX(t)=\vX_0$~: \emph{the center-of-mass becomes fixed}. Note that
while the frequency diverges, $\omega^*\to\infty$, the
radius of the circle, 
$\left|\big(eB(1-e\Theta B)\big)^{-1}\vcK/\right|$, shrinks to zero as the critical value is approached.
In fact, the \emph{whole system} gets ``frozen'' into a
static ``crystal'' configuration: 
 {all} {effective masses vanish},
$m_a^*=0$  when $B=B_c$; all individual positions are 
therefore fixed.
 
Eqn.
(\ref{exoNH})  only allows for $\vb=\const$~: the $4$-parameter
symmetry algebra reduces to mere translations
\footnote{All diffeomorphisms of the plane
become instead symmetries \cite{DHexo,winfty}.}.
 The associated conserved quantities behave as
 $\cP^i\to -eB_c\varepsilon^{ij}X_0^j$ and $\vcK\to0$
 as $M^*\to0$ i.e. $B\to B_c=(e\Theta)^{-1}$.

 Let us now restore the
electric field, $\vE$, assumed  constant for simplicity. This amounts to adding $\sum_ae_e\vE\cdot\vx_a$ to the Lagrangian (\ref{exoLag}). But the Kohn condition (\ref{exoKohnCond}) allows us to infer that this
is simply $e\vE\cdot\vX$;  the electric field only effects therefore the center-of-mass, but not the internal motion.

Let us now consider the center-of-mass equations,
(\ref{exoCMeqmot}). We show that the constant electric field can be eliminated as in the commutative case by a suitable \emph{Galilean boost}. Consider indeed 
$\vx_a\to\widetilde{\vx}_a=\vx_a+\vu t,\,
\vp_a\to\widetilde{\vp}_a=\vp_a+m_a\vu$,
where $\vu$ is a constant vector. It
 acts on the center-of-mass as
$
\vX\to\widetilde{\vX}=\vX+\vu t,
\;
\vP\to\widetilde{\vP}=\vP+M\vu.
$
This is not a symmetry of the system, though, but yields rather
\beqa
M^*\dot{\widetilde{X}^i}-\widetilde{P}^i=eM\Theta\big(Bu^i+\varepsilon^{ij}E^j\big),
\qquad
\dot{\widetilde{P}}^i-eB\varepsilon^{ij}\dot{\widetilde{X}^i}
=e\big(-B\varepsilon^{ij}u^j+E^i\big).
\nn
\eeqa
Choosing 
$
u^i=-\varepsilon^{ij}{E^j}/{B}
$
the electric term is therefore eliminated, leaving us with the
pure magnetic problem we just solved.
The motion in the original frame is thus obtained by
combining the purely-magnetic motions with the constant-speed drift $-\vu$, perpendicular to the electric field.
Off the critical case, $e\Theta B\neq1$, we get,  both for individual particles and their center-of-mass, collectively drifted cyclotronic motions with  common frequency $\omega^*$ plus drift velocity
$\varepsilon^{ij}{E^j}/{B}$.

In the critical case, $B=B_c$ and for $\vE=0$
instead, all ``motions'' are mere fixed points~: as we have seen, our particles are frozen into  static
equilibrium. Boosting backwards allows us to conclude that, when (\ref{critcase}) holds, \emph{all particles, and hence also their  center-of-mass, drift, collectively, with 
common Hall velocity} 
\beq
\dot{x}_a^i=\varepsilon^{ij}\frac{E^j}{B_c}
\qquad\hbox{for all}\quad
 a=1,\dots,N\,.
\label{Hallaw}
\eeq
By (\ref{inteqmot}), the internal motions are still frozen.

It is convenient to introduce the \emph{guiding centers} of the particles,
$ 
\vq_a=\vx_a-(m_a/e_aB_c^2)\vE
$ 
which move, together with their center of mass,
$ 
\vQ=(\sum_a{m_a\vq_a})/{M}=\vX-eM\Theta^2\vE,
$ 
following  the Hall law, (\ref{Hallaw}).
Their equations of motion can be derived
 from the reduced Lagrangian
[obtained from (\ref{exoLag}) by Faddeev-Jackiw reduction \cite{DHexo,NCLandau}],  which splits once again into c-o-m plus internal parts,
\beqa
L^{red}&=&%L^{red}_{com}+L^{red}_{int}=
\sum_a\frac{1}{2\theta_a}\varepsilon^{ij}q^i_a\dot{q}^j_a-\sum_a e_a\vE\cdot\vq_a\,
\label{reducedLag}
\\
&=&\left\{
\frac 1{2\Theta}\varepsilon^{ij}Q^i\dot{Q}^j-eE^iQ^i\right\}
+\sum_{a,b}
\frac{1}{4\Theta}
\frac{m_am_b}{M^2}\,\varepsilon^{ij}\left(q_a^i-q_b^i\right)
\left(\dot{q}_a^j-\dot{q}_b^j\right).
\nn
\eeqa

As noticed by Landau-Lifshitz in the commutative case, \cite{LaLi}, 
the guiding center coordinates  are \emph{non-commuting}, 
\beq
\{q^i_a,q^j_b\}=
-\theta_a\,\delta_{ab}\varepsilon^{ij}
=
-\frac{1}{e_aB_c}\delta_{ab}\varepsilon^{ij}\,,
\qquad
\{Q^i,Q^j\}=-\Theta\,\varepsilon^{ij}=
\frac{1}{eB_c}\varepsilon^{ij}.
\label{NCgc}
\eeq 
The Hamiltonian has no kinetic term and reduces the potential alone  \cite{Peierls}.  As 
\beq
 \delta L^{red}=\frac{\ d}{dt}
\Big(\sum_a
\frac{1}{2\theta_a}\varepsilon^{ij}q_a^j
-e_aE^it\Big)b^i=
\frac{d}{dt}\Big(\frac{1}{2\Theta}\varepsilon^{ij}Q^j-eE^it\Big)b_i
\eeq
 under a translation $\vb$,
the associated conserved quantities are non-commuting,
\beq
\cP^i_{red}=-\frac{1}{\Theta}\varepsilon^{ij}\Big(Q^j-\varepsilon^{jk}\frac{E^k}{B_c}\,t\Big),
\qquad
\{\cP^i_{red},\cP^j_{red}\}=-\frac{1}{\Theta}\,\varepsilon^{ij}=-eB_c\varepsilon^{ij},
\label{redconsmom}
\eeq
showing that our residual symmetry is the
\emph{Heisenberg group} with central extension parameter $-\Theta^{-1}=-eB_c$. These relations 
 are consistent with letting $B\to B_c$ in (\ref{exomagmomboost}) and (\ref{exoCMcomrel}).
Amusingly, the [Hall] motions can, once again,
be recovered from the conserved quantity,
\beq
Q^i(t)=\Theta\,\varepsilon^{ij}\cP^j_{red}+\varepsilon^{ij}\frac{E^j}{B_c}t.
\eeq

%%%%%%%%%%%%%%%%%%%%
\section{Conclusion}
%%%%%%%%%%%%%%%%%%%%

The intuitive meaning of the
Kohn condition is to guarantee 
 a collective behavior~: all particles
rotate with the same frequency, 
shared also by their center-of-mass. 
The additional condition $e_a\theta_a=\const$  implies
that the typical factors $(1-e_a\theta_a B)$ are the same for all particles, namely $(1-e\Theta B)$, 
allowing us to extend Kohn's theorem to exotic particles.

Our second result is to prove the 
  \emph{two-parameter centrally-extended ``\emph{exotic}''
Newton-Hooke symmetry} for our a system of $N$ exotic particles \footnote{Owing to the equivalence of the uniform-B--field and  oscillator
problems \cite{GiPo,ZH-Kohn}, the latter plainly shares the same exotic Newton-Hooke symmetry.}. 
As in the Galilean case, the commutation relations  only differ from the ordinary ($1$-parameter) case in the boost-boost relation,
which now also involves the non-commutative parameter $\Theta$,
and is supplemented by internal rotations and time translations.

Off the critical case $e\Theta B\neq1$, the motions are
analogous to those in the ordinary Landau problem but with modified frequency. In the critical case
$B=B_c=(e\Theta)^{-1}$, however, all particles are frozen into
a static ``crystal'' configuration when the electric field vanishes, and drift perpendicularly to the  electric field with  Hall velocity when $\vE=\const\neq0$.

It is worth saying that our description for the reduced system presented in Sect. \ref{Hall} is in fact that  
of \emph{pointlike vortices in the plane} \cite{NCLandau}, consistently with Laughlin's suggestion, who explains the Hall effect by the motion of charged vortices \cite{QHE}.

All our investigations have been purely classical. It is
not difficult to quantize our system, though, as in the ordinary case \cite{GiPo}.

%\newpage

\begin{acknowledgments} 
P.A.H is indebted to the \textit{
Institute of Modern Physics} of the Lanzhou branch of
the Chinese Academy of Sciences for hospitality, and to C. Duval for correspondence.
 This work was  partially supported by the National Natural Science Foundation of 
China (Grant No. 11035006) and by the Chinese Academy of Sciences visiting 
professorship for senior international scientists (Grant No. 2010TIJ06). 
\end{acknowledgments}
\goodbreak

%%%%%%%%%%%%%%%%%%%%%%%%%%%%%%%%%%%%%%%%%%%%%%%%%%%%%%%%%%%%%%%%%%%%
%%%%%%%%%%%%%%%%%%%%%%%%%%%%%%%%%%%%%%%%%%%%%%%%%%%%%%%%%%%%%%%%%%%%

\end{document}